# Why Do You Contribute to Stack Overflow? Understanding Cross-Cultural Motivations and Usage Patterns before the Age of LLMs

Sherlock A. Licorish
University of Otago
Dunedin, New Zealand
sherlock.licorish@otago.ac.nz

Elijah Zolduoarrati
University of Otago
Dunedin, New Zealand
zolduoarrati@gmail.com

Tony Savarimuthu
University of Otago
Dunedin, New Zealand
tony.savarimuthu@otago.ac.nz

Rashina Hoda
Monash University
Melbourne, Australia
rashina.hoda@monash.edu

Ronnie De Souza Santos
University of Calgary
Calgary, Canada
ronnie.desouzasantos@ucalgary.ca

Pankajeshwara Sharma
University of Fiji
Lautoka, Fiji
pankajeshwaras@unifiji.ac.fj

## ABSTRACT

**Background**: Understanding software practitioners' motivations for participating in community question-and-answer (CQA) platforms is crucial for sustaining knowledge-sharing ecosystems/portals (e.g., Stack Overflow), which is necessary to advance the discipline while also ensuring its longevity. This is particularly necessary in the age of LLMs, where data from such portals are used to train these models. **Objective**: However, limited insights exist regarding how contributors' motivations vary across different national cultures. This study investigates Stack Overflow contributors' motivations for platform participation and examines how these motivations differ across territories and relate to actual platform activities. **Method**: A mixed-methods approach was employed, combining qualitative content analysis of 600 "About Me" profiles with quantitative linguistic analysis of 268,215 contributors' data from the United States, China, and Russia. Using content analysis informed by established coding schemes, we identified 17 motivational categories. We utilized WordNet-based synonym extraction to analyze the complete dataset, followed by Spearman's correlation analysis to examine relationships between stated motivations and platform activities. **Results**: Findings show that contributors are primarily motivated by advertising opportunities and altruistic problem-solving desires. American contributors demonstrated stronger self-promotional behaviors while Chinese contributors exhibited greater learning-oriented engagement. Spearman's correlation analysis showed that those with more detailed profiles tend to engage in advertising and social activities, while learning-oriented users maintain minimal self-presentation. **Conclusion**: Understanding these motivational variations can inform strategies for enhancing cross-cultural participation in software engineering.

## KEYWORDS
Culture, Motivation, Participation, Stack Overflow

## 1 Introduction

Stack Overflow has served as a valuable knowledge source for practitioners since 2008 [1-4]. With the advent of Large Language Models (LLMs), a decline in contributors' activities has been reported since 2023 [50]. We believe it is timely to take stock of this trusted platform through the current disruptions caused by LLMs in software engineering (SE).

Software practitioners move from other platforms to Stack Overflow, where researchers identified three main motivating reasons [5] – availability of a larger developer community to source knowledge, availability of higher quality of Q&As and faster response time to obtain answers. Using semi-structured interviews, Coleman and Lieberman [2] identified seven motivations for Stack Overflow use: fun, ideology, reputation, reciprocity, efficacy, attachment, and structural capital. Subsequently, Abdalkareem, et al. [6] identified five developer motivations for using Stack Overflow based on analyzing their commits: benefiting from the use of existing knowledge, documenting bugs, promoting Stack Overflow (e.g., by adding tags), proposing feature or system improvements, and reusing available code. The use of existing knowledge was by far the largest category, accounting for 75% of the commits.

The study of Ndukwe, et al. [4] has identified four categories of practitioners' use of CQA websites such as Stack Overflow where they seek, gain, contribute and have quick access to knowledge. Of these four categories, three of them are about benefiting from the community and only one is about giving back to the community through individual contributions. Beyond discussing technical topics, developers are also motivated to discuss non-technical topics that are of interest to their careers such as sharing information about the current job market, workplace environment, and their career choices [7].

A systematic review conducted by Yang and Mao [1] has identified 134 unique motivations for the use of Stack Overflow from 41 related work, which are grouped into seventeen clusters. From these clusters they identified eight key factors that motivate participants to contribute. These are: obtaining a badge which signals their competence, improving their reputation score that signal both peer-recognition and competence, getting information, learning, reciprocity, helping others, obtaining privileges (which confers access to rare resources and encourages continuous contribution), and having fun.

These studies contribute to our understanding of how and why practitioners contribute to the open-source software (OSS) engineering community, which is necessary for the technological advancement of the discipline and its survival. However, limited insights exist on practitioners' reasons for contributing knowledge (knowledge-oriented motivations), and particularly across different cultures and geographies. Evidence suggests that a range of issues affect practitioners' willingness and interest in participating in Q&A communities, including their cultural background [8-10]. Thus, knowledge around the dynamics of online Q&A communities could be useful in helping to strategize around their long-term survival. This is particularly necessary in the age of LLMs, where data from community portals are used to train these models [51]. Evidence shows that when these models are trained on data that are self-generated (i.e., LLM responses to prompts) their performance declines [52]. Thus, diverse human solutions are central to advancing and training models to aid the community. We thus address this opportunity in this work by answering the following research questions (RQs).

**RQ1**. What are Stack Overflow contributors' motivations for participating on the platform?
**RQ2**. How do Stack Overflow contributors' motivations differ across territories/geographies?
**RQ3**. How are Stack Overflow contributors' motivations related to their actual activities on the platform?

Both RQ1 and RQ2 contribute unique insights to the community. RQ1 unearths motivational factors identifying latent universal patterns which provides a baseline understanding before conducting comparative analysis. RQ2 builds upon RQ1's findings by examining cultural variations in these established patterns. Also, despite both questions addressing the same unit of analysis (i.e., contributor motivations), they operate at different analytical levels; RQ1 focuses on factor identification, whereas RQ2 focuses on comparative analysis and cultural interpretation. RQ3 looks to explore and triangulate evidence of contributors' motivation with their actual activities on the platform, in providing insights into their behavior on CQA portal.

**Key contributions/statement of novelty** of this study include an understanding of the knowledge-oriented motivations of contributors on Stack Overflow as a representative Q&A community in directing strategies for their ongoing survival, as well as recommendations for practice and research. We examine contributors' motivations across multiple territories, contributing to the body of evidence aimed at understanding cultural differences and its impact on software engineering outcomes.

The remaining sections of this study are organized as follows. In the next section (Section 2) we review the literature and provide our methods in Section 3. Results are subsequently provided in Section 4. We discuss the findings and outline implications in Section 5, before considering threats to the study in Section 6. We provide concluding remarks and identify opportunities for future work in Section 7. We have uploaded a replication package [11] for those interested in examining our research process and outcomes[1], providing both raw and summarized data and details on the data extraction process.

## 2. Literature Review

To establish the current state of knowledge regarding existing scholarly work within this research domain, we conducted an informal literature review, allowing a more flexible approach to summarize the extant scholarly works. We first examine motivation in software engineering communities in Section 2.1. Next, we examine motivation driven by culture in software engineering communities in Section 2.2. Thereafter, we explore the literature on gender-based motivation in software engineering communities in Section 2.3, before examining personality-based motivation in software engineering communities in Section 2.4.

## 2.1. Motivation in Software Engineering Communities

Psychological theories of motivation have been applied and studied by researchers in the Software Engineering domain [12, 13]. Researchers have applied self-determination theory, in particular [12], and investigated the nature of motivations of practitioners by studying whether they are *intrinsic* or *extrinsic*, which focus on internal and external rewards respectively [1, 3]. Lu, et al. [3] have identified these two types of motivations in the context of acquiring badges. Intrinsic rewards include helping others and generating new knowledge, and extrinsic rewards include obtaining badges and higher reputation scores. Yang and Mao [1] have identified five different types of motivations, which they call regulation styles based on self-determination theory, which are: *external* (obtaining an external reward such as a badge), *introjected* (obtaining intangible rewards such as peer recognition), *identified* (where the goals pursued are recognized to be valuable to individuals such as getting information), *integrated* (e.g., where individuals believe the activities they pursue are valuable to them such as reciprocating knowledge sharing since the notion of 'community' is embodied through the practice of reciprocation) and *intrinsic* (obtaining internal personal rewards such as helping others and having fun).

---
[1] Remove to maintain blind review process.

Vadlamani and Baysal [14] have identified two categories of drivers that motivate the use of and participation on Stack Overflow – *personal* drivers and *professional* drivers. Personal drivers include helping others, having fun, self-learning and meeting their own needs. Professional drivers include improving skills and expertise, obtaining better answers and finding alternative solutions. These two categories (personal and professional drivers) can be further divided into the five regulation styles described above.

Unlike the above-mentioned works that mainly focus on motivating factors, Mustafa, et al. [15] considered both motivating and de-motivating factors (or barriers) for making consistent contributions to Stack Overflow, by applying psychological theory of social exchange. Based on hypotheses testing, their results show that Stack Overflow contributors who received positive votes, comments on their contributions, and downvotes are motivated to make active contributions to knowledge. On the other hand, those who mainly pose questions, who are recognized through upvotes, and those that receive negative votes are demotivated (or negatively influenced) to share knowledge. Researchers have not only identified motivations for active contributors who ask questions, post answers and comments, they have also considered lurkers who consume knowledge, but do not actively engage with knowledge creation [16, 17]. Lurkers' non-contributing behavior is attributed to their aversion to contributing to clutter, discomfort in interacting with strangers and being intimidated by the larger community size [17].

## 2.2. Culture-based Motivation in Software Engineering Communities

Cultural background of practitioners has been shown to motivate individuals differently [18-20]. The work of Oliveira, et al. [19] focused on identifying motivating and demotivating factors across three different cultures: American, Chinese and Indian. While the study reveals motivating factors such as efficiency in finding information and creating knowledge useful to others that are perceived across the three cultures, it also identified demotivating factors for certain cultures such as the inability to socialize and satisfy their curiosity when compared to other platforms such as Facebook. The study highlighted the mismatch in Stack Overflow design that encouraged mostly individualist values (i.e., most suitable for US participants) by focusing on efficient and to-the-point interactions between participants which may not be particularly suitable for users with mostly collectivist values (i.e., for Indian and Chinese participants). They posited that this misalignment could systematically hamper user engagement and contributions from collectivistic societies who value rich social interactions. Zolduoarrati, et al. [20] somewhat supported this position, noting that being sensitive to cultural background can enhance teamwork in global teams. Violation of human values in Stack Overflow such as hedonism and benevolence can also serve as demotivators for not engaging with the platform [18].

## 2.3. Gender-based Motivation in Software Engineering Communities

Another perspective previously considered when it comes to participation is gender. Differing gender-based motivations can be inferred from prior studies [8-10]. The work of Zolduoarrati and Licorish [10] noted that female participants use more collectivistic language and are more socially oriented than male participants, suggesting that awareness of their peculiar style and enabling a supporting environment may help their participation. In fact, it was shown that men tend to be motivated more by gamification than women [8]. Women are motivated to seek knowledge by asking more questions while men contribute to knowledge by providing more answers [9], providing complementary dynamics to the knowledge ecosystem. Lack of awareness in this regard could thus be threatening to women. For instance, Ford, et al. [17] have identified 14 barriers for female participation in Stack Overflow including fear of negative feedback and time constraints. Additionally, 11 barriers to social inclusion and diversity such as age, education, sexual orientation and race can demotivate practitioners from participating in Stack Overflow [21]. Eliminating such barriers can enhance practitioner participation.

**Table 1. Summary of identified motivation dimensions from related works**

| Dimension | Description | Components | Supporting References |
|---|---|---|---|
| Knowledge | Individual learning and community knowledge construction | Knowledge acquisition, contribution, sharing | [1, 3, 9, 14, 15, 19] |
| Social | Interpersonal interactions and reciprocity | Helping others, community reciprocation | [1, 3, 10, 14, 17, 19] |
| Signaling | Competence demonstration through platform metrics | Reputation scores, badge achievement | [1, 3, 22, 23] |
| Hedonistic | Entertainment and resource access through gamification | Fun, gamified participation | [8, 14, 18] |
| Diversity | Variation based on individual characteristics | Cultural, gender, personality differences | [8-10, 18-23] |

## 2.4. Personality-based Motivation in Software Engineering Communities

The personalities of practitioners may mediate their intrinsic motivations. Prior work has shown that authors with high reputation were extroverts when compared to users with low reputations [22]. Another study found authors with high reputation to be open-minded [23], suggesting that Stack Overflow usage motivations may also be influenced by personality types.

Table 1 summarizes different dimensions of motivation found in the literature. Seldom have prior work focused on investigating the role and impact of knowledge-oriented motivations across different cultures using a data-driven approach. Understanding how cultural differences influence knowledge-sharing behaviors is crucial for designing inclusive global platforms and fostering effective cross-cultural collaboration in software engineering. This is essential to support sustained knowledge ecosystems. This work bridges this gap by considering data from three different cultures. Thus, this paper is situated at the intersection of knowledge and cultural dimensions, aimed at unearthing insights into the motivations of diverse cultures in the knowledge creation and curation processes.

## 3. Study Settings

We present the data preparation, data analysis and measures for answering our RQs in this section. We first justify the dataset that was used for analysis and the procedures used for preparing the data in Section 3.1. Next, we document the qualitative and quantitative data analysis approaches that were used in Section 3.2, before providing the measures in Section 3.3.

### 3.1. Data Preparation

We reused the Stack Overflow dataset from Zolduoarrati et al. [20] to perform our investigation. In Zolduoarrati et al.'s study, it was noted that practitioners from United States of America (USA), China, and Russia were selected to explore the impact of individualism and collectivism cultural profiles on the behavior of Stack Overflow contributors. Contributors from these three countries were studied because they represent three of the largest user bases on Stack Overflow and exhibit distinct cultural dimensions on Hofstede's individualism-collectivism scale [20, 24], while also being in their respective regions [25]. Also, Russia is an interesting example of a high-income country with a moderate internet technology pace [26], and a useful benchmark for comparison against the USA and China given that they are competitors. Further, all three countries are advanced and highly developed, with a high proportion of internet users[2]. Stability in internet users accompanied by rapid technological development, render good illustrations around how these societies contribute to the global software engineering community. USA, China, and Russia have also invested in making the internet inclusive for all. Studying the data from these countries would help with understanding various contributors' motivations for participating in software engineering knowledge communities.

Contributors' countries of origin were detected by scanning their profiles using the Python library *Geotext*.[3] Although the tool is not able to detect the geographical location of all contributors due to a number of limitations (e.g., lack of profile, no mention of geographical location), the geographical spread it detects reflects the results of the 2019 Stack Overflow survey [27], and Zolduoarrati et al. [20] performed reliability evaluations of these records. The dataset extracted covered > 11 years (September 2008 to September 2019), as extracted from the Stack Exchange superset[4], comprising records for 268,215 contributors with complete profiles (USA = 222,162, China = 27,720 and Russia = 18,333).

Following Zolduoarrati et al. [20], we sought to extract all the quantitative footprints of developers' activities on the Stack Overflow platform. We extracted 11 dimensions mainly from the *Users* table, with join operations between the *Users* table and other tables from Stack Overflow based on their unique mutual identifiers (refer to Table 2). For instance, we use the *Posts* table to differentiate whether posts added by users are questions or answers. Given our drive to include all the possible measures, we anticipate that our work will be all encompassing, however, previous studies have examined several of these dimensions individually [24, 28-30]. Aggregate statistics for each dimension and each region may be seen in our replication package[5].

**Table 2. Descriptions of developers' Stack Overflow data**

| Dimension | Description |
| --- | --- |
| AboutMe + length | The number of words users wrote about themselves in their profiles' "About Me" sections. |
| Duration on site (months) | The number of months a user used the Stack Overflow platform. |
| Up Votes | A measure to reflect the usefulness of posts by a user. |
| Down Votes | A measure to indicate posts offering minimal value. |
| Reputation | A measure calculated by Stack Overflow to reflect the amount of trust the community has in a user. |
| Views | The total number of times a user profile has been viewed. |
| Badges | The total number of badges earned by a user. |
| Comments | The total number of comments provided by a user. |

---

[2] https://data.worldbank.org/indicator/IT.NET.USER.ZS

[3] https://github.com/elyase/geotext

[4] https://archive.org/details/stackexchange

[5] Replication package » Dimensions Aggregates.docx

| Dimension | Description |
| --- | --- |
| P questions | The total number of questions asked by a user. |
| P answers | The total number of answers provided by a user. |
| Post history edits | The total number of times a user has edited posts. |

## 3.2. Data Analysis

*Qualitative*: We adopted a directed content analysis approach seeded with codes from prior research, with inductive refinement where necessary to study contributors' motivation [31]. We began with a predefined coding scheme uncovered by an interview study of New Zealand practitioners that reused Stack Overflow data to seed our analysis [32], while remaining open to novel patterns that emerged during the analysis process. In this earlier study, practitioners identified nine reasons for consulting Stack Overflow: Search solutions, Find help, Debugging, Learning, Thinking, Familiarizing, Post questions, Post answers and comments, and Resource access. The qualitative data analysis was conducted using thematic analysis, adopting the approach by Braun and Clarke [33], to identify and interpret meaningful patterns. We explored contributors' Stack Overflow "About Me" profiles which typically introduces practitioners and their reason for joining the platform (refer to Figure 1 for example), and has been used to study various elements of practitioners' behaviors and intentions [10, 20].

Using the initial codes to seed our analysis (i.e., reasons for consulting Stack Overflow unearthed by [32]), we manually analyzed 600 "About Me" profiles from the dataset, representing 200 profiles from USA, 200 profiles from China and 200 profiles from Russia. For each cohort of "About Me" profiles, we randomly selected 100 profiles from the contributors with the highest reputation and 100 profiles from those with the lowest reputation in order to maintain a balanced view of the population of contributors. We then coded these 600 profiles. The procedure involved open coding where the "About Me" profiles were read and re-read for familiarization and if one of the nine aforementioned codes (motivation for using Stack Overflow) was observed, that initial code was assigned to the profile in question. These codes were explicit and clear in the surface level semantics of the data (see Braun and Clarke [33]).

Figure 1. Sample Stack Overflow profile

At times it was observed that the nine codes were not sufficient to capture the motivation of "About Me" profiles, where we then used thematic mapping to restructure the new codes (themes). Finally, following Braun and Clarke's [33] selective coding procedure, the resulting new codes were refined and organized into a coherent, internally consistent account, and a description was developed to accompany each new theme observed in the data. Eight new motivations emerged from the Stack Overflow data as shown in Table 3. When we could not clearly identify contributors' motivation in their "About Me" profile we labelled the profile "No Motivation".

As noted above, this study benefited from coding knowledge from previous coding efforts involving similar data and shared among several researchers [10, 20], however, independent reliability testing was performed in this study. A second coder independently coded 60 "About Me" profiles representing 10% of the coded data using the final coding scheme involving 17 codes. Thereafter, codes were compared where six codes differed among the two coders (i.e., 85% agreement).

Discussions then ensued, where there was consensus on the six codes (for three dimensions), resulting in 100% agreement. This knowledge was taken forward in another round of coding, where all the codes belonging to the three dimensions were re-examined for consistency.

*Quantitative*: In a second round of analysis, we extended the protocol/coding scheme to analyze all of the data (the entire dataset comprising 268,215 contributors' profiles) using a quantitative approach. First, we identified seed terms for the categories in Ndukwe et al. [32], and based on our coding process. We then retrieved all synonyms for these terms using an online thesaurus for the English language, WordNet, which is based on psycholinguistics studies [34]. WordNet has been used by several researchers carrying out linguistic analysis of Stack Overflow users (e.g., [35]). Care was taken to only include synonyms that were likely to be used in the context of descriptions of developers' motivation in their "About Me" profile on Stack Overflow. Finally, to increase the likelihood of capturing terms that match with contributors' motivation we extracted terms from the 600 "About Me" profiles for each of the profile coded based on the 17 categories. This approach to text mining has led to various insights around software development practitioners' behaviors (e.g., [36-39]). The coding scheme used for analyzing the data is provided in Table 3. Associated terms for each dimension may be seen in our replication package.

*Ethics statement*: We would like to acknowledge that all data used in this study are publicly available, anonymized, and analyzed in compliance with Stack Overflow's terms of service. We have also provided both raw and summarized data and details on the data extraction process as part of our replication package [11].

**Table 3. Coding scheme used, adapted from [32]**

| Dimension (number) | Description |
|---|---|
| Search solutions (1) | Find answers to programming problems. |
| Find help (2) | Get solutions for coding tasks. |
| Debugging (3) | Seek help when stuck debugging. |
| Learning (4) | Learn from examples on the platform. |
| Thinking (5) | Use feedback to develop thinking process. |
| Familiarizing (6) | Learn programming language syntax. |
| Post questions (7) | Ask questions in underexplored areas. |
| Post answers and comments (8) | Contribute by answering and commenting. |
| Resource access (9) | Quick access to programming resources. |
| Make friends (10) * | Connect with like-minded people. |
| Advertise themselves or Company (11) * | Showcase expertise for hiring opportunities. |
| Share ideas (12) * | Exchange ideas with other practitioners. |
| Increase reputation (13) * | Build platform reputation. |
| Wander (14) * | Browse the website casually. |
| Correct (15) * | Fix errors in others' posts. |
| Earn money directly (16) * | Monetize question answering. |
| Find jobs (17) * | Search for career opportunities. |

**Note**: * denotes new code from data

### 3.3. Measures for Answering RQs

To measure contributors' motivation (RQ1) we aggregated the codes that were recorded across the 17 categories in Table 3 for the 600 "About Me" profiles. We explored the differences in these codes and the patterns of outcomes to understand the reasons practitioners join the Stack Overflow portal. Follow up quantitative results were then extracted from the entire dataset using the linguistics patterns in Table 3, providing complementary outcomes. In measuring differences in contributors' motivation across territories (RQ2), we explored the codes and quantitative outcomes across the USA, China and Russia to analyze differences in the motivation of practitioners across these territories. Finally, in measuring the relation between contributors' motivation and platform activities (RQ3), we correlated the practitioners' motivation outcomes against their general activities on Stack Overflow. These latter analyses help us to assess practitioners' motivation against the way their activities are enacted during their time on the Stack Overflow platform.

## 4. Results

### RQ1. What are Stack Overflow contributors' motivation for participating on the platform?

Outcomes in Figure 2 depict the aggregated motivation for USA, Russia, and China combined. Here it is revealed that Stack Overflow practitioners use the portal to advertise the most, while many are driven by the desire to contribute and engage around problem solving on the platform. A large number of contributors have reported no motivation in their summary profile. Contributors also use Stack Overflow as a learning tool, where their questions and requests for help are addressed. Some contributors see the platform as a medium for making friends and socializing around problem solving, where other less notable reasons for contributors' engagement including increasing reputation, earning money and so on. Interestingly, much of the findings in Figure 2 (a) are triangulated by the quantitative linguistic findings in Figure 2 (b). A Chi-Squared test confirmed that differences are statistically significant (p < 0.05).

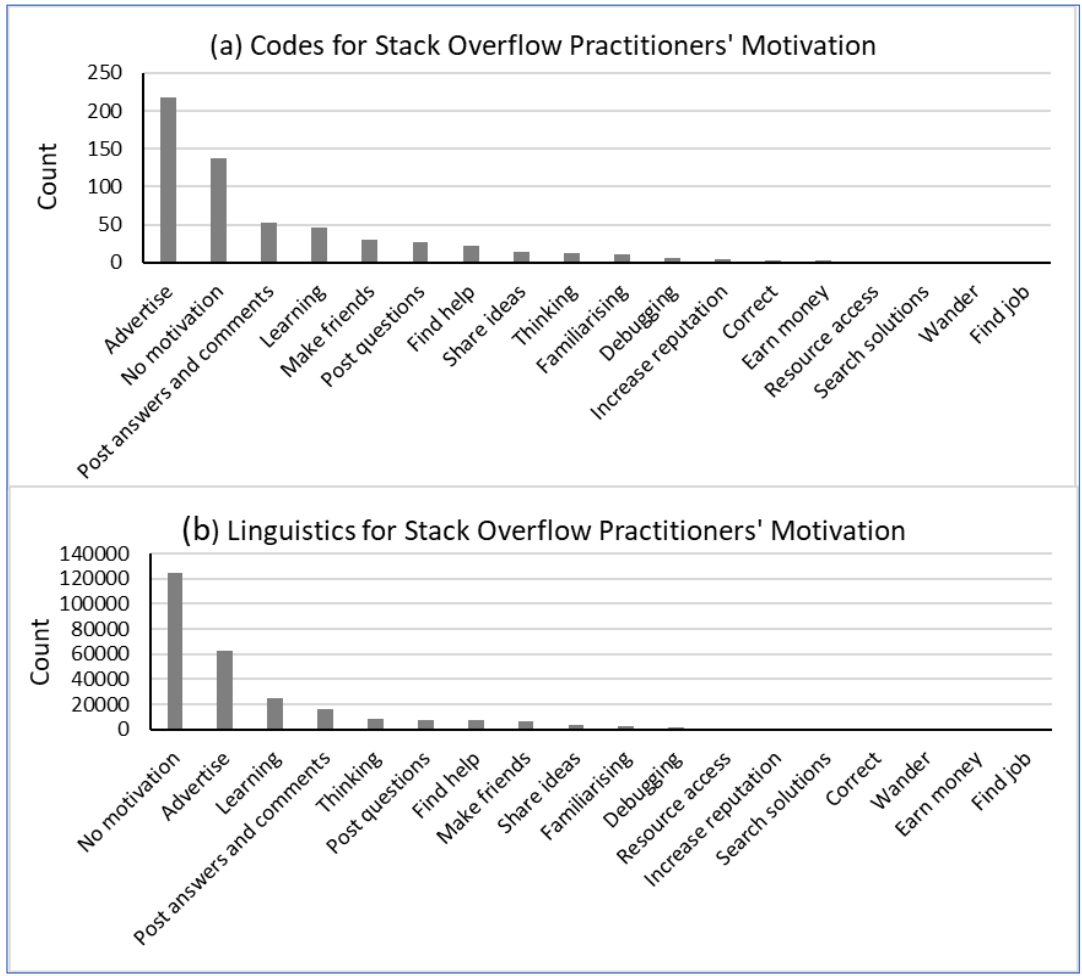

**Figure 2. Stack Overflow contributors' motivation**

### RQ2. How do Stack Overflow contributors' motivations differ across territories?

We visualize the differences in Stack Overflow contributors' motivation across USA, China and Russia in Figure 3. While some members across all territories did not express motivation or were focused on posting answers and comments or making friends, practitioners in the USA used Stack Overflow for advertisement much more that those of China and Russia. On the contrary, this pattern of outcome was reversed for learning, where contributors from China used Stack Overflow for learning more than twice as much as those from USA and Russia. We observed a slight divergence in the results of the qualitative and quantitative outcomes in Figure 3 (a) and Figure 3 (b), where some of the terms were not captured in the linguistic patterns mined.

### RQ3. How are Stack Overflow contributors' motivations related to their actual activities on the platform?

We present Spearman's correlation results in Table 4, where Stack Overflow contributors' motivation was correlated with their activities as captured by various metrics from the platform. We explore noteworthy statistically significant correlations (r >= 0.3). It is observed that the length of the "About Me" correlated with contributors' motivation to advertise, meaning that when contributors spent time adding detailed "About Me" text on Stack Overflow, they tend to be driven by a need to advertise. The length of contributors' "About Me" profiles showed significant positive correlations with motivation to share ideas, develop reputation, find friends, post answers and comments, and engage in thinking and problem solving. Conversely, contributors with detailed "About Me" profiles demonstrated significantly lower motivation for learning. Contributors who were primarily motivated by problem-solving activities also tended to have shorter engagement durations.

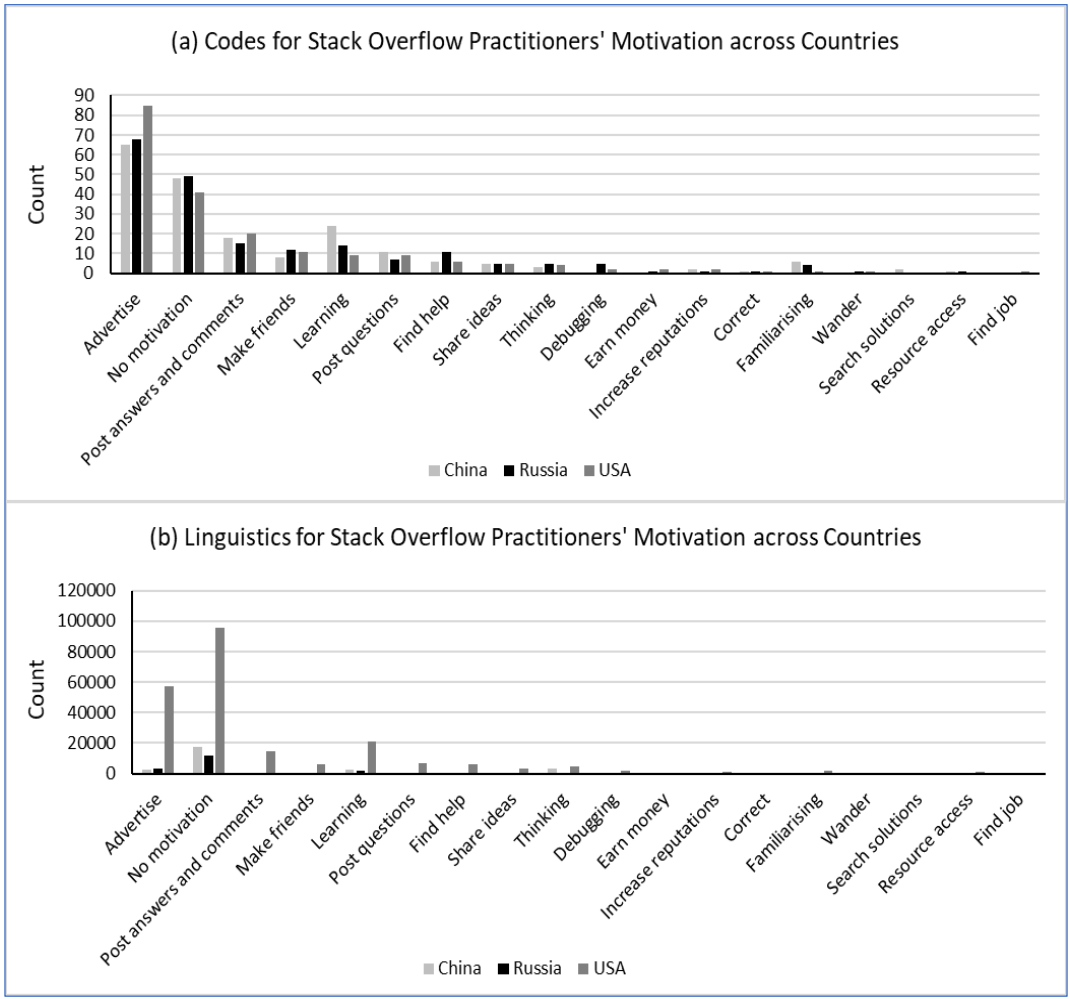

Figure 3. Stack Overflow contributors' motivations across territories

Table 4. Relations between contributors' motivations and platform activities

| Dimension | AboutMe | Duration | Up Votes | Down Votes | Reputation | Views |
|---|---|---|---|---|---|---|
| **SearchSolutions** | **-0.005** | **-0.008** | -0.002 | -0.001 | -0.002 | -0.001 |
| **FindHelp** | **-0.010** | **-0.020** | -0.003 | 0.000 | 0.001 | 0.000 |
| **Debugging** | 0.000 | **-0.012** | 0.003 | -0.001 | 0.000 | -0.001 |
| **Learning** | **-0.100** | **-0.020** | **-0.013** | **-0.005** | **-0.012** | **-0.006** |
| **Thinking** | **0.100** | **-0.100** | **-0.014** | -0.003 | **-0.010** | **-0.005** |
| **Familiarizing** | **-0.011** | **0.005** | 0.000 | -0.001 | -0.003 | -0.001 |
| **PostQuestions** | **0.023** | **-0.007** | -0.001 | 0.003 | 0.003 | 0.002 |
| **PostAnswersAndComments** | **0.107** | **-0.040** | **0.021** | **0.011** | **0.015** | **0.009** |
| **ResourceAccess** | **0.030** | **-0.010** | 0.000 | -0.001 | -0.003 | -0.002 |
| **findFriends** | **0.100** | **-0.024** | **0.020** | **0.004** | **0.013** | **0.009** |
| **advertise** | **0.350** | **-0.020** | **0.005** | **0.005** | **0.012** | **0.012** |
| **shareIdeas** | **0.200** | **0.033** | **-0.010** | 0.001 | 0.000 | -0.001 |
| **reputations** | **0.160** | **-0.04** | -0.002 | -0.001 | 0.001 | -0.001 |
| **wander** | **0.024** | 0.000 | 0.003 | **0.013** | **0.007** | 0.003 |
| **correct** | **0.070** | **-0.012** | **0.009** | **0.023** | **0.005** | **0.007** |
| **earnMoney** | **0.021** | -0.002 | **0.009** | **0.005** | **0.017** | **0.006** |
| | Badges | Comments | Questions | Answers | History | |
| **SearchSolutions** | -0.003 | -0.001 | -0.002 | -0.002 | -0.001 | |
| **FindHelp** | -0.001 | 0.003 | -0.001 | 0.002 | 0.001 | |
| **Debugging** | -0.001 | -0.001 | -0.003 | 0.000 | 0.000 | |
| **Learning** | **-0.012** | **-0.009** | 0.002 | **-0.011** | **-0.008** | |

| | | | | | |
|---|---|---|---|---|---|
| Thinking | -0.022 | -0.006 | -0.016 | -0.007 | -0.008 |
| Familiarizing | -0.003 | -0.001 | -0.004 | -0.001 | -0.001 |
| PostQuestions | 0.003 | 0.003 | **0.005** | 0.002 | 0.002 |
| PostAnswersAndComments | **0.014** | **0.018** | 0.005 | **0.016** | **0.017** |
| ResourceAccess | -0.002 | -0.001 | 0.001 | -0.001 | -0.001 |
| findFriends | **0.012** | **0.008** | 0.002 | **0.010** | **0.017** |
| advertise | **0.010** | **0.012** | 0.000 | **0.010** | **0.008** |
| shareIdeas | -0.007 | -0.001 | **-0.005** | -0.002 | -0.002 |
| reputations | **-0.006** | -0.001 | **-0.006** | 0.001 | 0.000 |
| wander | **0.006** | **0.004** | 0.001 | **0.005** | 0.003 |
| correct | 0.003 | **0.006** | -0.003 | 0.002 | **0.009** |
| earnMoney | **0.012** | **0.007** | 0.004 | **0.009** | **0.010** |

Bold values = p < 0.05.

## 5. Discussion and Implications

### RQ1. What are Stack Overflow contributors' motivations for participating on the platform?

For RQ1, findings reveal that Stack Overflow contributors are primarily motivated by advertising opportunities and altruistic desires to contribute to problem-solving activities, yet a notable proportion of users did not explicitly state their motivations. This aligns with prior research on online community participation, where both self-promotional and prosocial motivations coexist [40, 41]. The fact that advertising appears as a strong declared motive suggests users are also leveraging the platform for self-presentation, consistent with the platform's integration into career advancement strategies [42]. That said, this could be troublesome if the interests of specific technology groups are the sole focus of those contributing on the portal, especially given that data from community portals are used to train the recently more popular AI-solutions (and LLMs). Ideally, knowledge on community portals should be wide-ranging, and representative of a wide breadth of diverse views.

The triangulation between qualitative and quantitative findings strengthens the validity of our findings, where users stated intentions align with their linguistic expressions in their About Me (see Figure 2). The prevalence of learning-oriented participation supports the concept of legitimate peripheral participation, to which newcomers engage to acquire expertise while gradually increasing their contributions [43]. This should be safeguarded, in supporting the survival of diverse open software engineering communities.

### RQ2. How do Stack Overflow contributors' motivations differ across territories?

For RQ2, American contributors tend to demonstrate higher advertising motivations compared to those of Chinese and Russian. Chinese contributors exhibited greater learning-oriented participation. These findings may somewhat be linked towards Hofstede's individualism-collectivism framework, where Americans' self-promotional behaviors can signal higher individualistic values [44]. The elevated learning motivation among Chinese users may reflect the rapid technological development and knowledge acquisition imperatives within China's evolving software industry, where knowledge-seeking behaviors can be readily seen in emerging technology markets [45]. In a similar vein, American developers' focus on advertising aligns with a mature job market where personal branding and visibility are crucial for career advancement. Inf fact, the Chinese contributors' emphasis on learning may indicate a developing market with substantial skill acquisition needs [46]. On the other hand, the lower advertising motivation in Russia and China may suggest different norms where direct self-promotion through technical platforms is less culturally accepted [47]. These findings are revealing, in that it is not anticipated that a community portal reserved for problem solving would illustrate such distinct cultural differences, given that software engineering and technology in general is held to be a global language. Outcomes here show that even when not intentional, cultural and other behaviors become apparent, and so, provision should be made for this on community portals, especially in limiting the potential negative effects on open and widespread participation.

### RQ3. How are Stack Overflow contributors' motivations related to their actual activities on the platform?

For RQ3, the correlation analysis reveals particularly strong relationships involving profile elaboration ("About Me" length) and various motivational dimensions which mostly aligns with existing research. For one, the positive correlation between "About Me" length and **advertising** motivation (r = 0.350) provides validation that users who invest time in detailed self-presentation are indeed driven by promotional goals, supporting signaling theory in online environments [48]. Interesting findings encompass the negative correlations between "About Me" and **learning** (r = -0.100), combined with the positive correlation towards **thinking** (r = 0.100) but negative correlation with duration on the website (r = -0.100). Findings here seem to suggest that learning-oriented users maintain minimal profiles and instead focus on content consumption as their goal is not self-interest, while problem-solving enthusiasts invest in profile development but engage in shorter, more focused sessions in achieving their objective. These findings align with research on expert-novice differences in information-seeking strategies [49]. Again, motives that are not aligned with community development and openness could derail community values aimed at collective knowledge development. In this regard, platform moderators on these portals have an imperative to encourage and support unbiased participation.

## 6. Threats to Validity

First, threats to internal validity concerns the reliability of *Geotext* Python library to identify contributors' countries of origin from their profiles. This automated approach may introduce systematic errors, as contributors might not accurately represent their actual geographical location in their profiles, or relocate after profile creation without updating their information. Additionally, the manual coding process of "About Me" profiles, despite achieving 100% agreement after discussion between coders, introduces researcher bias in interpretation. The selection of high and low reputation contributors (100 each per country) for manual analysis may not represent the broader population of Stack Overflow users, and that inferences may not represent typical user behavior.

Threats to external validity is also limited by the focus on only three countries (USA, China, and Russia), which may not represent the diversity of global software development communities or cultural contexts. In fact, the exclusion of contributors without clear geographical indicators may systematically exclude certain user populations, even those from either USA, China, or Russia. Next, the study period spanning September 2008 to September 2019 does not reflect current motivational patterns, particularly considering the rapid evolution of online platforms and changing developer demographics. Further, Stack Overflow represents only one type of CQA platform, such that findings may not necessarily propagate to other technical communities with different governance structures or user interfaces.

Threats to construct validity involves the operationalization of motivation through "About Me" profiles, which may not capture the full spectrum of user motivations or may reflect socially desirable responses (i.e., self-reporting bias) rather than genuine motivations. In other words, contributors may not accurately articulate their true motivations in their own profiles. Linguistic analysis approach using WordNet synonyms may miss culturally-nuanced motivational statements. On the other hand, the aggregation of 17 motivational categories into broader patterns may oversimplify motivational structures that could be more complex than meets the eye. Additionally, the correlation analysis between stated motivations and platform activities assumes that behavioral metrics accurately reflect underlying motivational drives, when alternative explanations such as platform constraints, skill levels, or time availability could account for fluctuations in usage activity.

## 7. Conclusion and Future Work

This study investigated the motivations driving practitioners' participation in Stack Overflow across different cultural contexts, specifically examining contributors from the USA, China, and Russia. Through a mixed-methods approach combining content analysis of 600 "About Me" profiles and quantitative linguistic analysis using WordNet of data from 268,215 contributors, we sought to understand both the nature of motivational factors and their cultural variations. Our methodology involved directed content analysis using established coding schemes, followed by Spearman's correlation analysis to examine relationships between stated motivations and actual platform activities.

Stack Overflow contributors are primarily driven by advertising opportunities and altruistic problem-solving desires, yet a large proportion did not state any motivations in their profiles. The triangulation between qualitative and quantitative analyses strengthens these conclusions, demonstrating consistency between contributors' stated intentions and their linguistic expressions. Importantly, cultural differences emerged as significant factors influencing participation patterns, with American contributors showing stronger self-promotional behaviors while Chinese contributors demonstrated greater learning-oriented engagement. These patterns align with established cultural frameworks, while also reflecting different technological maturity levels across three regions. The correlation analysis between motivational dimensions and platform activities highlighted that contributors with detailed profiles tend to engage in advertising and social connection activities, while learning-oriented users maintain minimal self-presentation and focus on content consumption. The negative correlation between learning motivation and profile elaboration, combined with shorter engagement durations for problem-solving enthusiasts, suggests distinct user archetypes with different platform utilization patterns.

Future work may expand the geographical scope to include more diverse regions and sub-cultures. Longitudinal analysis can also be worthwhile to monitor temporal changes in contributors' motivations, as platforms and developer ecosystems evolve. Such an avenue would provide deeper insights into how engagement drivers change due to many drivers. Additionally, comparative studies across different CQA platforms, such as Reddit or Quora, or even across Stack Exchange sites, could reveal how platform design features influence motivational structures. In the age of LLMs where data on community portals form the source of knowledge to train such models, sustaining participation from diverse communities have significant implications for sustaining representative knowledge ecosystems.

# REFERENCES

[1] Y. Yang and X. Mao, "Understanding Developers' Contribution Motivation in Stack Overflow: A Systematic Review," in *2023 30th Asia-Pacific Software Engineering Conference (APSEC)*, 2023: IEEE, pp. 359-368, doi: 10.1109/APSEC60848.2023.00046.

[2] E. Coleman and Z. Lieberman, "Contributor motivation in online knowledge sharing communities with reputation management systems," in *Proceedings of the 2015 Annual Research Conference on South African Institute of Computer Scientists and Information Technologists*, 2015, pp. 1-12, doi: 10.1145/2815782.2815810.

[3] Y. Lu et al., "Motivation under gamification: An empirical study of developers' motivations and contributions in Stack Overflow," *IEEE transactions on software engineering*, vol. 48, no. 12, pp. 4947-4963, 2021, doi: 10.1109/TSE.2021.3130088.

[4] I. G. Ndukwe, S. A. Licorish, and S. G. MacDonell, "Perceptions on the utility of community question and answer websites like Stack Overflow to software developers," *IEEE Transactions on Software Engineering*, vol. 49, no. 4, pp. 2413-2425, 2022, doi: 10.1109/TSE.2022.3220236.

[5] M. Squire, ""Should We Move to Stack Overflow?" Measuring the Utility of Social Media for Developer Support," in *2015 IEEE/ACM 37th IEEE International Conference on Software Engineering*, 2015, vol. 2: IEEE, pp. 219-228, doi: 10.1109/ICSE.2015.150.

[6] R. Abdalkareem, E. Shihab, and J. Rilling, "What do developers use the crowd for? a study using Stack Overflow," *IEEE Software*, vol. 34, no. 2, pp. 53-60, 2017, doi: 10.1109/MS.2017.31.

[7] A. Barua, S. W. Thomas, and A. E. Hassan, "What are developers talking about? an analysis of topics and trends in Stack Overflow," *Empirical software engineering*, vol. 19, pp. 619-654, 2014, doi: 10.1007/s10664-012-9231-y.

[8] H. Cavusoglu, Z. Li, and K.-W. Huang, "Can gamification motivate voluntary contributions? The case of StackOverflow Q&A community," in *Proceedings of the 18th ACM Conference Companion on Computer Supported Cooperative Work & Social Computing*, 2015, pp. 171-174, doi: 10.1145/2685553.2698999.

[9] Y. Wang, "Understanding the reputation differences between women and men on Stack Overflow," in *2018 25th Asia-Pacific Software Engineering Conference (APSEC)*, 2018: IEEE, pp. 436-444, doi: 10.1109/APSEC.2018.00058.

[10] E. Zolduoarrati and S. A. Licorish, "On the value of encouraging gender tolerance and inclusiveness in software engineering communities," *Information and Software Technology*, vol. 139, p. 106667, 2021, doi: 10.1016/j.infsof.2021.106667.

[11] Anonymous. On Understanding Practitioners' Motivation for Participating on Stack Overflow, Zenodo, doi: 10.5281/zenodo.17156137.

[12] R. M. Ryan and E. L. Deci, "Self-determination theory and the facilitation of intrinsic motivation, social development, and well-being," *American Psychologist*, vol. 55, no. 1, p. 68, 2000, doi: 10.1037//0003-066x.55.1.68

[13] C. França, F. Q. Da Silva, and H. Sharp, "Motivation and satisfaction of software engineers," *IEEE Transactions on Software Engineering*, vol. 46, no. 2, pp. 118-140, 2018, doi: 10.1109/TSE.2018.2842201.

[14] S. L. Vadlamani and O. Baysal, "Studying software developer expertise and contributions in Stack Overflow and GitHub," in *2020 IEEE International Conference on Software Maintenance and Evolution (ICSME)*, 2020: IEEE, pp. 312-323, doi: 10.1109/ICSME46990.2020.00038.

[15] S. Mustafa, W. Zhang, and M. M. Naveed, "What motivates online community contributors to contribute consistently? A case study on Stackoverflow netizens," *Current Psychology*, vol. 42, no. 13, pp. 10468-10481, 2023, doi: 10.1007/s12144-022-03307-4.

[16] A. E. Abbas, "Investigating 'One-Day Flies' Users in The StackOverflow: Why Do and Don't People Participate?," in *2019 International Conference on ICT for Smart Society (ICISS)*, 2019, vol. 7: IEEE, pp. 1-5, doi: 10.1109/ICISS48059.2019.8969815.

[17] D. Ford, J. Smith, P. J. Guo, and C. Parnin, "Paradise unplugged: Identifying barriers for female participation on Stack Overflow," in *Proceedings of the 2016 24th ACM SIGSOFT International Symposium on Foundations of Software Engineering*, 2016, pp. 846-857, doi: 10.1145/2950290.2950331.

[18] S. Krishtul et al., "Human Values Violations in Stack Overflow: An Exploratory Study," in *Proceedings of the 26th International Conference on Evaluation and Assessment in Software Engineering*, 2022, pp. 70-79, doi: 10.1145/3530019.3530027.

[19] N. Oliveira, M. Muller, N. Andrade, and K. Reinecke, "The exchange in StackExchange: Divergences between Stack Overflow and its culturally diverse participants," *Proceedings of the ACM on Human-computer Interaction*, vol. 2, no. CSCW, pp. 1-22, 2018, doi: 10.1145/3274399.

[20] E. Zolduoarrati, S. A. Licorish, and N. Stanger, "Impact of individualism and collectivism cultural profiles on the behaviour of software developers: A study of stack overflow," *Journal of Systems and Software*, vol. 192, p. 111427, 2022/10/01/ 2022, doi: 10.1016/j.jss.2022.111427.

[21] B. T. R. Savarimuthu, Z. Zareen, J. Cheriyan, M. Yasir, and M. Galster, "Barriers for social inclusion in online software engineering communities-a study of offensive language use in gitter projects," in *Proceedings of the 27th International Conference on Evaluation and Assessment in Software Engineering*, 2023, pp. 217-222, doi: 10.1145/3593434.3593463.

[22] B. Bazelli, A. Hindle, and E. Stroulia, "On the personality traits of StackOverflow users," in *2013 IEEE international conference on software maintenance*, 2013: IEEE, pp. 460-463, doi: 10.1109/ICSM.2013.72.

[23] M. Papoutsoglou, G. M. Kapitsaki, and L. Angelis, "Modeling the effect of the badges gamification mechanism on personality traits of Stack Overflow users," *Simulation Modelling Practice and Theory*, vol. 105, p. 102157, 2020, doi: 10.1016/j.simpat.2020.102157.

[24] E. Zolduoarrati, S. A. Licorish, and N. Stanger, "Harmonising Contributions: Exploring Diversity in Software Engineering through CQA Mining on Stack Overflow," *ACM Transactions on Software Engineering and Methodology* 2024, doi: 10.1145/3672453.

[25] D. W. Jorgenson and K. Vu, "Information technology and the world economy," *Scandinavian Journal of Economics*, vol. 107, no. 4, pp. 631-650, 2005, doi: 10.1111/j.1467-9442.2005.00430.x.

[26] International Telecommunication Union. "ICT Facts and Figures [Fact sheet]." https://www.itu.int/en/ITU-D/Statistics/Documents/facts/ICTFactsFigures2017.pdf. (accessed 4 August, 2025).

[27] T. Bachschi, A. Hannak, F. Lemmerich, and J. Wachs, "From Asking to Answering: Getting More Involved on Stack Overflow," *arXiv preprint arXiv:.04025*, 2020, doi: 10.48550/arXiv.2010.04025.

[28] S. A. Licorish and M. Wagner, "Dissecting copy/delete/replace/swap mutations: insights from a GIN case study," presented at the Proceedings of the Genetic and Evolutionary Computation Conference Companion, Boston, MA, 2022. doi: 10.1145/3520304.3533970

[29] O. P. Omondiagbe, S. A. Licorish, and S. G. Macdonell, "Evaluating Simple and Complex Models' Performance When Predicting Accepted Answers on Stack Overflow," in *2022 48th Euromicro Conference on Software Engineering and Advanced Applications (SEAA)*, 31 Aug.-2 Sept. 2022 2022, pp. 29-38, doi: 10.1109/SEAA56994.2022.00014.

[30] A. Tahir, J. Dietrich, S. Counsell, S. Licorish, and A. Yamashita, "A large scale study on how developers discuss code smells and anti-pattern in Stack Exchange sites," *Information and Software Technology*, vol. 125, p. 106333, 2020/09/01/ 2020, doi: 10.1016/j.infsof.2020.106333.

[31] S. A. Licorish and S. G. MacDonell, "Understanding the attitudes, knowledge sharing behaviors and task performance of core developers: A longitudinal study," *Information and Software Technology*, vol. 56, no. 12, pp. 1578-1596, 2014/12/01/ 2014, doi: 10.1016/j.infsof.2014.02.004.

[32] I. G. Ndukwe, S. A. Licorish, and S. G. MacDonell, "Perceptions on the Utility of Community Question and Answer Websites Like Stack Overflow to Software Developers," *IEEE Transactions on Software Engineering*, vol. 49, no. 4, pp. 2413-2425, 2023, doi: 10.1109/TSE.2022.3220236.

[33] V. Braun and V. Clarke, "Using thematic analysis in psychology," *Qualitative Research in Psychology*, vol. 3, no. 2, pp. 77-101, 2006/01/01 2006, doi: 10.1191/1478088706qp063oa.

[34] G. A. Miller, "WordNet: a lexical database for English," *Commun. ACM*, vol. 38, no. 11, pp. 39–41, 1995, doi: 10.1145/219717.219748.

[35] L. Tóth, B. Nagy, T. Gyimóthy, and L. Vidács, "Mining Hypernyms Semantic Relations from Stack Overflow," presented at the Proceedings of the IEEE/ACM 42nd International Conference on Software Engineering Workshops, Seoul, Republic of Korea, 2020. doi: 10.1145/3387940.3392160

[36] S. A. Licorish and S. G. MacDonell, "What affects team behavior? Preliminary linguistic analysis of communications in the Jazz repository," in *2012 5th International Workshop on Co-operative and Human Aspects of Software Engineering (CHASE)*, 2-2 June 2012 2012, pp. 83-89, doi: 10.1109/CHASE.2012.6223029.

[37] S. A. Licorish and S. G. MacDonell, "The true role of active communicators: an empirical study of Jazz core developers," presented at the Proceedings of the 17th International Conference on Evaluation and Assessment in Software Engineering, Porto de Galinhas, Brazil, 2013. doi: 10.1145/2460999.2461034

[38] S. A. Licorish and S. G. MacDonell, "Differences in Jazz project leaders' competencies and behaviors: A preliminary empirical investigation," in *2013 6th International Workshop on Cooperative and Human Aspects of Software Engineering (CHASE)*, 25-25 May 2013 2013, pp. 1-8, doi: 10.1109/CHASE.2013.6614725.

[39] S. A. Licorish and S. G. MacDonell, "Communication and personality profiles of global software developers," *Information and Software Technology*, vol. 64, pp. 113-131, 2015/08/01/ 2015, doi: 10.1016/j.infsof.2015.02.004.

[40] M. M. Wasko and S. Faraj, "Why should I share? Examining social capital and knowledge contribution in electronic networks of practice," *MIS Quarterly,* vol. 29, no. 1, pp. 35–57, 2005, doi: 10.5555/2017245.2017249.

[41] O. Nov, M. Naaman, and C. Ye, "Analysis of participation in an online photo-sharing community: A multidimensional perspective," *Journal of the American Society for Information Science and Technology,* vol. 61, no. 3, pp. 555-566, 2010, doi: 10.1002/asi.21278.

[42] S. K. Kuttal, X. Chen, Z. Wang, S. Balali, and A. Sarma, "Visual Resume: Exploring developers' online contributions for hiring," *Information and Software Technology,* vol. 138, p. 106633, 2021/10/01/ 2021, doi: 10.1016/j.infsof.2021.106633.

[43] C. Filstad and J. McManus, "Transforming knowledge to knowing at work: the experiences of newcomers," *International Journal of Lifelong Education,* vol. 30, no. 6, pp. 763-780, 2011/12/01 2011, doi: 10.1080/02601370.2011.625573.

[44] G. Hofstede, Culture's Consequences: International Differences in Work-Related Values, 1st ed. Thousand Oaks, CA: SAGE, 1984.

[45] N. K. Park, J. M. Mezias, J. Lee, and J.-H. Han, "Reverse knowledge diffusion: Competitive dynamics and the knowledge seeking behavior of Korean high-tech firms," *Asia Pacific Journal of Management,* vol. 31, no. 2, pp. 355-375, 2014/06/01 2014, doi: 10.1007/s10490-013-9349-5.

[46] O. Victor Barinua and V.-J. Olatokunbo, "The impact of skill acquisition on entrepreneur development," *Saudi Journal of Business and Management Studies,* vol. 7, no. 5, pp. 137-146, 2022, doi: 10.36348/sjbms.2022.v07i05.004.

[47] E. Meyer, *The Culture Map*. New York City, NY: PublicAffairs, 2014.

[48] K. A. Basoglu and T. J. Hess, "Online Business Reporting: A Signaling Theory Perspective," *Journal of Information Systems,* vol. 28, no. 2, pp. 67-101, 2014, doi: 10.2308/isys-50780.

[49] M. M. L. Chiu, S. K. W. Chu, K. K. K. Ting, and G. Y. C. Yau, "A novice-expert comparison in information search," *CITE Research Symposium, CITERS 2011,* 2012 2011. [Online]. Available: http://hdl.handle.net/10722/161204.

[50] R. M. del Rio-Chanona, N. Laurentsyeva, and J. Wachs, "Large language models reduce public knowledge sharing on online Q&A platforms," *PNAS nexus*, 3(9), pp. 400, 2024.

[51] A. Al-Kaswan and M. Izadi, "The (ab) use of open source code to train large language models," In 2023 IEEE/ACM 2nd International Workshop on Natural Language-Based Software Engineering (NLBSE) (pp. 9-10). IEEE, 2023.

[52] I. Shumailov, Z. Shumaylov, Y. Zhao, N. Papernot, R. Anderson, and Y. Gal, "AI models collapse when trained on recursively generated data," *Nature*, 631(8022), pp.755-759, 2024.